%% file: main.tex
\pdfoutput=1 
\documentclass[%
reprint,
superscriptaddress,
nobibnotes,
nolongbibliography,
amsmath,amssymb,
aps,
prl,
floatfix,
]{revtex4-2}
\usepackage{packages, packages_figs}

\begin{document}
\title{Quantum gate dynamics beyond the rotating-wave approximation using multi-timescale quantum averaging theory}

\author{Kristian D. Barajas}

%
%
%
%
%

\affiliation{%
 Department of Physics \& Astronomy, University of California -- Los Angeles, California 90095, USA
}

\affiliation{%
Mani L. Bhaumik Institute for Theoretical Physics, University of California, Los Angeles, California 90095, USA 
}

\affiliation{%
Center for Quantum Science and Engineering, University of California -- Los Angeles, California 90095, USA
}

\affiliation{
NSF Challenge Institute for Quantum Computation, University of California -- Berkeley, California 94720, USA}

\author{Wesley C. Campbell}%

\affiliation{%
 Department of Physics \& Astronomy, University of California -- Los Angeles, California 90095, USA
}

\affiliation{%
Center for Quantum Science and Engineering, University of California -- Los Angeles, California 90095, USA
}

\affiliation{
NSF Challenge Institute for Quantum Computation, University of California -- Berkeley, California 94720, USA}

\date{\today}
\vspace{1cm}

\begin{abstract}
\input{sections/I_Abstract}
\end{abstract}
   \maketitle

\input{sections/II_Intro}

\input{sections/III_QAT}

\input{sections/IV_Examples}

\input{sections/V_Conclusion}

  

\bibliography{main.bib}

\input{supplemental_arxiv}

\end{document}

%% file: sections/I_Abstract.tex
We present a quantum averaging theory (QAT) for analytically modeling unitary gate dynamics in driven quantum systems beyond the rotating-wave approximation.
QAT addresses the simultaneous presence of distinct timescales by generating a rotating frame with a dynamical phase operator that toggles with the high-frequency dynamics and yields an effective Hamiltonian for the slow degree of freedom. 
By accounting for the fast-varying effects, we demonstrate that high-fidelity two-qubit gates in strongly driven systems are achievable by going beyond the validity of first-order approximations. 
The QAT results rapidly converge with numerical calculations of a fast-entangling M{\o}lmer-S{\o}rensen trapped-ion-qubit gate in the strong coupling regime, illustrating QAT's ability to simultaneously provide both an intuitive, effective-Hamiltonian model and high accuracy. 

%% file: sections/II_Intro.tex
Quantum logic gates, the fundamental operations of quantum computation, perform operations on qubits akin to Boolean logic gates on classical bits \cite{nielsenQuantumComputationQuantum2010}. 
However, physical realizations involve driving quantum systems over extended times where accurately modeling the unitary dynamics presents significant analytic and numerical challenges due to the presence of multiple timescales.

For example, hardware-level interactions used to effect a basic $X$ or $Y$ gate can frequently be mapped onto a two-level system resonantly interacting with an AC electromagnetic field \cite{sakuraiModernQuantumMechanics2020}.
Under the rotating-wave approximation (RWA) \cite{footAtomicPhysics2005} the AC field effectively induces coherent population inversion, which is essential for executing basic quantum logic gates.
Without the RWA this seemingly simple example lacks an exact solution due to the presence of distinct timescales: the slow Rabi oscillations overlaid with fast beat note dynamics, leading to small gate errors.
These errors are further compounded in systems with multiple near- and off-resonant interactions such as stimulated Raman transitions \cite{bergmannPerspectiveStimulatedRaman2015}, multi-level qudit configurations \cite{omanakuttanQuantumOptimalControl2021,lowControlReadout13level2025}, and qubit coupling via bosonic modes in trapped-ion or circuit QED architectures \cite{sorensenEntanglementQuantumComputation2000,blaisCavityQuantumElectrodynamics2004,majerCouplingSuperconductingQubits2007}.
Achieving fault tolerance will likely require a level of precision where off-resonant effects cannot be overlooked.

Several advanced techniques exist to isolate the slowly-varying dynamics, including high-frequency averaging \cite{schererNewPerturbationAlgorithms1997,jamesEffectiveHamiltonianTheory2007,jauslinQuantumAveragingDriven2000,casasFloquetTheoryExponential2001,rahavEffectiveHamiltoniansPeriodically2003,zeuchExactRotatingWave2020,bukovUniversalHighFrequencyBehavior2015,eckardtHighfrequencyApproximationPeriodically2015,goldmanPeriodicallyDrivenQuantum2014}, Fer-type successive frame changing \cite{ferResolutionLequationMatricielle1958,madhuFerExpansionEffective2006,manangaApplicationFloquetMagnusFer2019,sametiStrongcouplingQuantumLogic2021}, transitionless quantum driving \cite{berryTransitionlessQuantumDriving2009, petiziolFastAdiabaticEvolution2018}, and projection-based methods \cite{schriefferRelationAndersonKondo1966,sanzAdiabaticEliminationEffective2016,paulischAdiabaticEliminationHierarchy2014,tevrugtMoriZwanzigProjectionOperator2019,casasTimedependentPerturbationTheory2015, arnalExponentialPerturbativeExpansions2020}. 
Building on prior quantum averaging methods \cite{kamelPerturbationMethodTheory1970,caryLieTransformPerturbation1981,frascaUnifiedTheoryQuantum1992,buitelaarMethodAveragingBanach1993,schererQuantumAveragingIII1998}, Quantum Averaging Theory (QAT) provides a unified, unitary framework that systematically separates fast and slow degrees of freedom across multiple timescales with minimal assumptions about system structure. 
This separation is achieved through partial averaging, which acts as a low-pass filter, generating a rotating frame that isolates the slowly varying interactions in an effective Hamiltonian.

Unlike Floquet-based expansions \cite{goldmanTheoreticalAspectsHigherorder1992,levanteFormalizedQuantumMechanical1995,casasFloquetTheoryExponential2001,rahavEffectiveHamiltoniansPeriodically2003,manangaFloquetMagnusExpansion2015,goldmanPeriodicallyDrivenQuantum2014,eckardtHighfrequencyApproximationPeriodically2015} requiring periodic, far-detuned drives or multi-mode methods \cite{ernstDecouplingRecouplingUsing2005, casasTimedependentPerturbationTheory2015, arnalExponentialPerturbativeExpansions2020} such as the widely used Effective Hamiltonian Theory (EHT) \cite{jamesEffectiveHamiltonianTheory2007} with predefined resonance conditions, QAT employs an inductive, order-by-order procedure to isolate near-resonant contributions while retaining off-resonant effects, typically discarded in the RWA, for improved analytic accuracy.
This provides a generalizable, algorithmic strategy that remains effective for quasi-periodic, non-commensurate, and rapidly modulated systems, beyond the RWA limit where other approaches break down.
Here ``beyond RWA'' refers not only to counter-rotating terms but more generally to off-resonant contributions that are typically neglected under RWA assumptions.
Moreover, the approach is valid on infinite-dimensional Hilbert spaces \cite{buitelaarMethodAveragingBanach1993}, particularly important for architectures using Fock space operators as qubits \cite{kochChargeinsensitiveQubitDesign2007,devoretSuperconductingCircuitsQuantum2013,leghtasStabilizingBellState2013,ofekExtendingLifetimeQuantum2016} or a quantum bus \cite{majerCouplingSuperconductingQubits2007,blaisCircuitQuantumElectrodynamics2021,sorensenQuantumComputationIons1999,sorensenEntanglementQuantumComputation2000}.

In this Letter, we apply the QAT framework to enable high-fidelity quantum gates beyond the RWA limit in systems driven across multiple timescales.
This architecture-independent approach provides accurate fidelity estimation and informs strategies for error suppression for a variety of gate schemes as demonstrated on the Mølmer–Sørensen gate \cite{sorensenEntanglementQuantumComputation2000}.
As shown through comparisons with full numerical simulations, QAT converges rapidly, even in regimes where previous methods typically struggle or fail.
For full derivations and generalizations, we refer the reader to the companion paper \cite{barajasQuantumAveragingTheory2025}.

%% file: sections/III_QAT.tex
\section{The QAT Framework}\label{sec:perturbation}
In a suitable interaction picture, we assume a bounded interaction Hamiltonian depending on a small parameter $0\le \lambda< 1$ \cite{sakuraiModernQuantumMechanics2020}, $\hat{H}_I(s;\lambda) = \sum_{n=1}^{\infty} \lambda^n \hat{H}^{(n)}_I(s)$.
We further assume $\hat{H}_I(s;\lambda)$ is sufficiently well-behaved to admit an almost-periodic Fourier series expansion,
\begin{equation}\label{eq:qat:Fourier_modes}
	\hat{H}_I ^{(n)}(s) = \hat{H}_{I,0}^{(n)} + \sum_{\mathclap{k=1}}^{\mathcal{N}(n)} ( e^{i s\Lambda ^{(n)}_{\omega_k}  }\,\, \hat{h}^{(n)}_{I,k} + h.c. )
\end{equation}
with $\mathcal{N}(n)$ total (dimensionless) base frequencies $\Lambda ^{(n)}_{\omega_k} \doteq \omega^{(n)}_k / \omega_0$, scaled by the characteristic frequency of the unperturbed system, $\omega_0 \sim \lVert \hat{H}_0 \rVert/\hbar$ \footnote{The relevant energy scale for arbitrary $\hat{H}_0$ is typically given by its spectral norm $\lVert \hat{H}_0 \rVert$ (\textit{i.e.}, the largest singular value).}.
This structure allows $\hat{H}_I(s;\lambda)$ to characterize interactions on multiple timescales (\textit{i.e.}\ almost-periodic) where Floquet-based techniques typically fail. 
Slow dynamics arise when a base frequency or beat frequency, arising from higher-order interactions, approaches resonance. 
The (near-) resonant interactions dominate the long-time behavior, making them a natural candidate for inclusion in an effective Hamiltonian description.


In the QAT framework, the interaction propagator is assumed to be separable into the QAT-factorized form
\begin{empheq}[box=\tightfbox]{equation}\label{eq:qat:interaction_propagator_factorized}
	\begin{aligned}
		\hat{U}_I(s;\lambda) &= \hat{U}_\mathrm{fast}(s;\lambda) \, \hat{U}_\mathrm{eff}(\tau;\lambda) \big\rvert_{\tau=\lambda s}
	\end{aligned}
\end{empheq}
where the fast propagator $\hat{U}_\mathrm{fast}(s;\lambda)$ characterizes the fast-varying modulations on the slowly-varying envelope of the effective propagator $\hat{U}_\mathrm{eff}(\tau;\lambda)$.
To achieve this dynamical separation of timescales the aim is to construct an effective Hamiltonian $\hat{H}_{I,\mathrm{eff}}(\tau;\lambda) = \sum_{n=1}^{\infty} \lambda^n \, \hat{H}^{(n)}_{I,\mathrm{eff}}(\tau)$ parameterized by a lagging ``time'' variable $\tau \doteq \lambda s$, characterizing the long-time $s \gtrsim \mathcal{O}(1/\lambda)$ interactions free from fast-varying effects.
The effective Hamiltonian will govern an ``effective'' interaction picture equation 
\begin{equation}\label{eq:qat:average_eq}
    i \lambda \, \partial_\tau \hat{U}_\mathrm{eff}(\tau;\lambda) = \hat{H}_{I,\mathrm{eff}}(\tau;\lambda) \, \hat{U}_\mathrm{eff}(\tau;\lambda)
\end{equation}
yielding only the slow-time ($\tau$) dynamics of $\hat{H}_I(s;\lambda)$. 
Meanwhile, $\hat{U}_\mathrm{fast}(s;\lambda)$ captures the influence that rapidly-varying $s \lesssim \mathcal{O}(1)$ interactions typically dropped under RWA assumptions have on the long-time dynamics.

To perturbatively generate the effective Hamiltonian we proceed with an asymptotic expansion.
As in prior Magnus-based methods \cite{casasFloquetTheoryExponential2001,casasTimedependentPerturbationTheory2015,arnalExponentialPerturbativeExpansions2020,eckardtHighfrequencyApproximationPeriodically2015}, we use an exponential Lie transformation 
\begin{equation}\label{eq:qat:exp_Lie_transformation}
    \hat{U}_\mathrm{fast}(s;\lambda) = e^{-i \hat{\Phi}(s; \lambda)}, \quad \hat{\Phi}(s;\lambda) = \sum_{n=1}^{\infty} \lambda^n \, \hat{\Phi}^{(n)}(s)
\end{equation}
depending on a \textit{dynamical phase} operator $\hat{\Phi}(s;\lambda)$ \cite{magnusExponentialSolutionDifferential1954}. 
An immediate advantage over a time-ordered Dyson series is preservation of unitarity (enforcing the preservation of Hermiticity) upon truncation and providing a simple inverse property $(e^{i \hat{\Phi}})^{-1} = e^{-i \hat{\Phi}}$ \cite{blanesMagnusExpansionIts2009}.


Transforming $\hat{H}_I(s;\lambda)$ with respect to \cref{eq:qat:interaction_propagator_factorized,eq:qat:exp_Lie_transformation} yields a Magnus homological equation of the form
\begin{empheq}[box=\tightfbox]{equation}\label{eq:qat:QAT_homological}
            \partial_s \hat{\Phi}^{(n)}= \hat{\mathcal{H}}^{(n)}_{\Phi}(s) - \hat{H}_{I,\mathrm{eff}}^{(n)}(\tau)
\end{empheq}
where the auxiliary Hamiltonian operator is defined as
\begin{equation}\label{eq:appendix:QAT_aux_generator}
\boxed{
	\hat{\mathcal{H}}_{\Phi}^{(n)} = \hat{H}_I^{(n)} + \sum_{k=1}^{n-1} \frac{B_k}{k!} \, \left((-)^k \,  \hat{S}_{k}^{(n)} - \hat{T}_{k}^{(n)} \right)
    }
\end{equation}
where $B_k$ are Bernoulli numbers.
The operators $\hat{S}_k^{(n)}$ and $\hat{T}_k^{(n)}$ are generated by recurrence relations
\begin{subequations}\label{eq:appendix:homological_recurrence}
    \begin{align}
	    \hat{S}_{0}^{(n)} &= \hat{H}_{I}^{(n)}, \quad
     \hat{T}_{0}^{(n)} = \hat{H}_{I,\mathrm{eff}}^{(n)} \\
	    \hat{A}_{k}^{(n)} &= \sum_{m=1}^{n-k} \left[ i \hat{\Phi}^{(m)},\hat{A}_{k-1}^{(n-m)} \right], \quad 1 \le k \le n-1
    \end{align}
\end{subequations}
where $\hat{A}$ is to be replaced by $\hat{S}$ or $\hat{T}$ and the adjoint action $\mathrm{ad}_{\hat{X}}(\hat{Y}) = [\hat{X},\hat{Y}]$ and $\mathrm{ad}_{\hat{X}}^{(k)}(\hat{Y}) = [\hat{X},\mathrm{ad}_{\hat{X}}^{(k-1)}(\hat{Y})]$ for integer $k \ge 2$.
The first three orders are provided for the reader's convenience with explicit time dependence omitted:
\begin{equation}\label{eq:appendix:aux_op_table}
\begin{split}
	\hat{\mathcal{H}}_{\Phi}^{(1)} &= \hat{H}_I^{(1)} \\
	\hat{\mathcal{H}}_{\Phi}^{(2)} &= \hat{H}_I ^{(2)} + \frac{1}{2} [i \hat{\Phi}^{(1)}, \hat{H}_I ^{(1)} + \hat{H}{\vphantom{H}}^{(1)}_{I,\mathrm{eff}}] \\
	\hat{\mathcal{H}}_{\Phi}^{(3)} &= \hat{H}_I^{(3)} + \frac{1}{2} \left( [i \hat{\Phi} ^{(2)}, \hat{H}_I^{(1)} + \hat{H}{\vphantom{H}}^{(1)}_{I,\mathrm{eff}}] \right. \\
        &\quad \left. + [i \hat{\Phi} ^{(1)}, \hat{H}_I^{(2)} +\hat{H}{\vphantom{H}}^{(2)}_{I,\mathrm{eff}}] \right) \\
        &\quad - \frac{1}{12} \left( [i \hat{\Phi} ^{(1)},[i \hat{\Phi}^{(1)},\hat{H}{\vphantom{H}}^{(1)}_{I,\mathrm{eff}}-\hat{H}_I^{(1)}]] \right ).
\end{split}
\end{equation} 
Once Eq.~\eqref{eq:qat:QAT_homological} is solved, the integration constant can be ignored to remove any $s_0$ dependence, ensuring a manifestly gauge-invariant expansion \cite{eckardtHighfrequencyApproximationPeriodically2015,barajasQuantumAveragingTheory2025}.

We employ a partitioned expansion by timescale separation (PETS) approach, in which, at each order, $\hat{\mathcal{H}}^{(n)}_{\Phi}(s)$ is separated into a fast-varying $\hat{\mathcal{H}}^{(n)}_{\Phi,>}(s)$, slow-varying $\hat{\mathcal{H}}^{(n)}_{\Phi,<}(s \! = \! \tau/\lambda)$, and constant $\hat{\mathcal{H}}^{(n)}_{\Phi,0}$ contribution for a high-frequency cutoff $\lambda_\mathrm{cutoff} = \lambda$ \footnote{More generally, the frequency cutoff must obey $\lambda \le \lambda_\mathrm{cutoff} <1$ with improved results for $\lambda \le \lambda_\mathrm{cutoff} \ll 1$ \cite{barajasQuantumAveragingTheory2025}}.
For a gauge-invariant and asymptotically valid QAT expansion, we require Eq.~\eqref{eq:qat:QAT_homological} be regularized by demanding that
\begin{subequations}\label{eq:qat:QAT_choice}
\begin{empheq}[box=\fbox]{align}
            \overline{\hat{\Phi}^{(n)}(s)} ={}& 0 \; \Leftrightarrow \; \hat{\Phi}^{(n)}(s) = \int^s \! ds' \; \hat{\mathcal{H}}^{(n)}_{\Phi,>}(s') \label{eq:qat:QAT_phase_cond} \\
	    \hat{H}^{(n)}_{I,\mathrm{eff}}(\tau) ={}& \overline{\hat{\mathcal{H}}{\vphantom{H}}^{(n)}_{\Phi}(s)} = \hat{\mathcal{H}}{\vphantom{H}}^{(n)}_{\Phi,0} +  \hat{\mathcal{H}}{\vphantom{H}}^{(n)}_{\Phi,<}(s\!=\!\tau/\lambda) \label{eq:qat:QAT_effHam_cond}
\end{empheq}
\end{subequations}
where we use the ``partial'' time averaging procedure 
\begin{equation}\label{eq:qat:time_average_twotime}
	\overline{\hat{A}(s)} = \hat{A}_0 + \hat{A}_<(s) \cong \int_{-\infty}^\infty \!ds' \: \hat{A}(s') \: f(s-s')     
\end{equation}
which removes fast-varying effects by effectively applying the RWA, \emph{i.e.} an idealized low-pass filter $f(s)$ with a cutoff frequency $\lambda$.
The regularizing conditions ensure $\hat{\Phi}(s;\lambda)$ reproduces the fast-varying dynamics and $\hat{H}_{I,\mathrm{eff}}(\tau;\lambda)$ describes the slowly-varying interaction from $\hat{H}_I(s;\lambda)$.
While the PETS approach shares qualities with EHT \cite{jamesEffectiveHamiltonianTheory2007}, the QAT formalism doesn't suffer from non-unitary and non-Hermitian artifacts, is valid to desired order of approximation, and retains the effects of off-resonant contributions required for accurate modeling.

Repeating the process to arbitrary $N$th-order, $\hat{U}_I(s;\lambda)$ is approximated by the truncated QAT solution
\begin{equation}\label{eq:qat:QAT_interaction_approx}
	\hat{U}^{[N]}_I(s;\lambda) = \hat{U}_\mathrm{fast}^{[N-1]}(s;\lambda) \hat{U}^{[N]}_\mathrm{eff}(\tau;\lambda) \Big\rvert_{\tau=\lambda s}
\end{equation}
for the time-evolution of the interaction picture state $\ket{\psi_I^{[N]}(s;\lambda)} = \hat{U}^{[N]}_I(s;\lambda) \hat{U}^{\dagger [N]}_I(s_0;\lambda) \ket{\psi(s_0)}$ where $\hat{U}_\mathrm{fast}^{[N-1]}(s;\lambda) = \exp(-i \hat{\Phi}^{[N-1]}(s;\lambda))$, and $\hat{U}^{[N]}_{\mathrm{eff}}(\tau;\lambda)$ is the exact or perturbative (up to $\mathcal{O}(\lambda^N)$) solution to $\hat{H}^{[N]}_{\mathrm{eff}}(\tau;\lambda)$. 
Error bounds are discussed in \cite{barajasQuantumAveragingTheory2025}.


%% file: sections/IV_Examples.tex
\section{Example: Fast-Entangling M{\o}lmer-S{\o}rensen Quantum Gate}\label{sec:MS_gate}

\input{sections/fig_MS_level_diagram.tex}
\input{sections/fig_MS_Gate.tex}

We apply the QAT framework to generate a high-fidelity quantum logic gate in a trapped-ion architecture consisting of ion qubits with atomic resonance frequency $\omega_{eg}$ allowed to oscillate along a single direction with one motional mode at secular frequency ${\nu}$. 

Consider a traveling standing-wave arrangement \cite{jamesQuantumDynamicsCold1998,haffnerQuantumComputingTrapped2008,sanerBreakingEntanglingGate2023} of the fast-entangling operation proposed by M{\o}lmer and S{\o}rensen \cite{sorensenEntanglementQuantumComputation2000}, using counter-propagating, global laser fields ($l=B,R$) with a relative $\pi$ phase offset: 
\begin{equation}
\hat{V}_l(t) = \frac{\Omega}{2} (-1)^{\delta_{l,R}}\Bigl(\hat{J}_{+} e^{i\eta_l (\hat{a}^\dagger + \hat{a})}\, e^{-i( \omega_l t - \phi_l)} + h.c. \Bigr)
\end{equation}
where $\hat{J}_{\pm} = \sum_i \hat{\sigma}_\pm^{(i)}$ are the total qubit raising and lowering operators;  $\eta_{B/R}=\pm \eta$ is the Lamb-Dicke parameter; $\hat{a}$ ($\hat{a}^\dagger$) is the phonon annihilation (creation) operator; and $\omega_{B/R} = \omega_{eg} \pm \Delta$ are equally blue- ($+$) and red-detuned ($-$) from the qubit resonance as seen in Fig.~\eqref{fig:level_diagrams}
\footnote{For simplicity, we have disregarded the optical counter-rotating terms $\propto 2 \omega_{eg}$ appearing in the interaction picture.
Leading order corrections are of strength $\mathcal{O}(\Omega/2\omega_{eg}) \ll 1$, which for $\omega_{eg} \gg \nu \gg \Omega$ are significantly smaller than the perturbative regime probed.}.

We focus on the strong-coupling regime $\eta\Omega/\nu \sim 0.1$ where terms typically neglected in the RWA and first-order Lamb-Dicke approximation become significant. 
Using QAT, we simultaneously track both (near-)resonant and off-resonant interactions and incorporate their corrections into the QAT approximation.
We define $s = {\nu} t$, $\lambda = \eta$, and $\Lambda_\omega \doteq \omega/{\nu}$ ($\Lambda_{\nu} = 1$) such that in the interaction picture of the ions and the motional mode \footnote{The system is sequentially transformed, first with respect to $\hat{H}_{0,\omega_{eg}}$ and then $\hat{H}_{0,{\nu}}$, then scaled by the secular frequency.}, 
\begin{equation}\label{eq:ex:MS_Interaction_Hamiltonian}
\hat{H}_I(s) ={} \Lambda_{\Omega}'e^{\eta^2/2}\hat{J}_{\phi,y}\Bigl(f(s) \hat{\mathcal{D}}(\alpha(s)) +h.c. \Bigl)
\end{equation}
with $f(s)= ie^{i\phi_{-}}e^{-i\Lambda_{\Delta} s}$, $\Lambda_\Omega' = e^{-\eta^2/2}\Lambda_\Omega$, $\mathcal{\hat{D}}(\alpha) = e^{\alpha \hat{a}^\dagger - \alpha^* \hat{a}}$, $\alpha(s) = i \eta e^{i s\Lambda_{\nu}}$, $\phi_{\pm} = (\phi_B \pm \phi_R)/2$, and $\hat{J}_{\phi,y} = \sin(\phi_+)\hat{J}_{x}+\cos(\phi_+) \hat{J}_{y}$ for total Pauli spin operators $\hat{J}_k = \frac{1}{2} \sum_{i}\sigma_k^{(i)}$.
Expanding in powers of $\lambda=\eta$,
\begin{equation}\label{eq:ex:MS_Interaction_Hamiltonian_nth}
\hat{H}_I^{(n)}\!=i^n \Lambda_\Omega' \hat{J}_{\phi,y} \mathcal{\hat{D}}^{(n)}(\sfrac{\alpha(s)}{\lambda}) \!
\begin{cases}
    \mathrm{Re}[f(s)], & n \; \mathrm{even} \\
    i\mathrm{Im}[f(s)] , & n \; \mathrm{odd}
\end{cases}
\end{equation}
where $\mathcal{\hat{D}}^{(n)}(\alpha)$ is the $n$th-order Taylor polynomial of the normal-ordered displacement operator (see supplemental material).
In the Lamb-Dicke regime we recover the familiar first motional sidebands ($n=1$) with outside regime effects appearing for $n\ge2$.
A valid perturbative treatment requires transforming into an interaction picture with respect to the $n=0$ carrier term.
Since the carrier commutes with $\hat{H}_I(s)$ at all times, it does not affect higher-order dynamics; we therefore neglect it in the following and consider its effect separately.

In the fast entanglement protocol \cite{sorensenEntanglementQuantumComputation2000}, $\Delta$ is tuned near the first motional sideband with a small detuning $\Lambda_{\delta}=\nu -\Lambda_{\Delta} =\lambda \Lambda_{\epsilon}  \leq \lambda$.
Defining $G(\tau) = e^{i\Lambda_{\epsilon} \tau}e^{i\phi_{-}}$, from Eq.~\eqref{eq:qat:QAT_choice} and \eqref{eq:ex:MS_Interaction_Hamiltonian_nth}, the first-order QAT results yield
\begin{subequations}
\begin{align}
    \partial_s\hat{\Phi}^{[1]}(s) =& -\eta\Lambda_{\Omega}' \hat{J}_{\phi,y} (a^\dagger G^*(s)e^{2i\Lambda_{\nu}s} \!+h.c.)\label{eq:ex:MS_dynamical_phase} \\
    \hat{H}^{[1]}_{I,\mathrm{eff}}(\tau) =& - \eta \Lambda_{\Omega}'  \hat{J}_{\phi,y} (\hat{a}^\dagger G(\tau) + h.c.) \label{eq:ex:MSgate_effective_Hamiltonian}
\end{align}
\end{subequations}
where the latter admits the well-known solution \cite{sorensenEntanglementQuantumComputation2000}
\begin{equation}\label{eq:ex:MS_Gate_eff_o2}
		\hat{U}_\mathrm{eff}^{[1]}(\tau,\tau_0=0;\lambda) ={} \mathcal{\hat{D}}\left(\hat{J}_{\phi,y} \alpha_{\mathrm{ms}}(\tau)  \right) e^{i \vartheta(\tau) \hat{J}_{\phi,y}^2}
\end{equation}
with displacement $\alpha_{\mathrm{ms}}(\tau) = i\Lambda_{\Omega}' \{ G\}_\tau$ and geometric phase $\vartheta(\tau) = \mathrm{Im}[\{ \alpha^*_{\mathrm{ms}} \partial_{\tau}\alpha_{\mathrm{ms}}\}_\tau]=\Lambda_\Omega^{\prime 2}\mathrm{Im}[\{\{G^*\}_\tau G\}_\tau]$, where we define $\{ f\}_s\equiv \int_0^s ds' f(s')$.
The maximally-entangled Bell state $\ket{\boldsymbol{\varphi}_{+}} = (\ket{ee} + \ket{gg})/\sqrt{2}$ (up to a global phase) is ideally generated at the gate time $s_g$ if $\hat{U}_{I}(s_g) = \exp(i \frac{\pi}{2} \hat{J}_{\phi,y}^2)$ with $\phi_{+}=\pi/4$.
To lowest-order accuracy, a coherent gate is achieved when $\vartheta(\tau_g\!=\!\lambda s_g) = \pi/2$, aligned with round trips in $xp$-phase space where $\alpha_{\mathrm{ms}}(\tau_g) = 0$, largely decoupling the vibrational motion from the internal states \cite{sorensenEntanglementQuantumComputation2000}.
However, in strong driving regimes, the first-order QAT interaction propagator, $\hat{U}_I^{[1]}(s;\lambda) = \hat{U}_{\mathrm{eff}}^{[1]}(\tau;\lambda)\rvert_{\tau=\lambda s}$, as shown in Fig.~\eqref{fig:MSgate_Eff}, underestimates the significant effects of off-resonant terms and out-of-Lamb-Dicke processes, providing an insufficient approximation of the gate operation.

To more accurately describe the gate dynamics, corrections to the effective Hamiltonian arising from off-resonant processes, particularly the counter-rotating anti-Jaynes-Cummings terms, must be included.
QAT systematically accounts for all such contributions to any desired order; a complete listing of results up to fourth order is given in the Appendix.
For example, the full second-order QAT result yields
\begin{subequations}
\begin{align}
   \lambda\hat{\Phi}^{(1)}(s) \;& 
    \begin{aligned}[t]
    ={}& \frac{1}{2} \hat{J}_{\phi,y} (i \alpha_{\mathrm{cr}}(s)a^\dagger+h.c.) \\
    \end{aligned} \\
    \lambda^2\hat{H}_{I,\mathrm{eff}}^{(2)}(s) \;&
    \begin{aligned}[t]
    ={}& \hat{J}_{\phi,y}^2 \mathrm{Im}[\overline{\alpha_{\mathrm{cr}}^{*}(s)\partial_s\alpha_{\mathrm{cr}}(s)}] \\
    \end{aligned}
\end{align}
\end{subequations}
where 
\begin{equation}\label{eq:ex:alpha_cr_secondorder}
\begin{aligned}
\alpha_{\mathrm{cr}}(s) ={}& i\eta\int^s ds'\Lambda_{\Omega}' G^*(s') e^{2i \Lambda_{\nu} s'} \\
\end{aligned}
\end{equation}
introducing a small correction to the geometric phase from the self-interaction of the counter-rotating terms.
This result provides analytic insight into the pulse profiles required to suppress the leading order off-resonant interactions at the gate time. 
For example, applying a smooth window function that terminates at the gate time effectively decouples the system from both the carrier and the first-order counter-rotating terms.

We also consider subleading fidelity losses due to corrections in the effective Hamiltonian beyond the first-order Lamb-Dicke regime, where the gate becomes sensitive to these effects \cite{sorensenEntanglementQuantumComputation2000,sametiStrongcouplingQuantumLogic2021}.
Focusing first on the dominant contributions and neglecting off-resonant effects and multi-photon corrections, the effective dynamics are governed predominately by single-photon processes $\hat{H}_{I,\mathrm{eff}}(\tau) \approx \overline{\hat{H}_I}(\tau;\lambda)= \Lambda_\Omega \hat{J}_{\phi,y} ( i\mathcal{\hat{D}}_1(\eta) G(\tau) + h.c.)$ where ${\mathcal{\hat{D}}}_{1}(\eta) = e^{-\eta^2/2}\sum_{k \ge 0} \frac{(i \eta)^{2k+1}}{(k+1)!k!} \hat{a}^\dagger \hat{a}^{\dagger k} \, \hat{a}^k$, exciting the first motional sidebands via net phonon absorption or emission.
Single-photon effects are already suppressed to all orders under the same condition requiring $\alpha_\mathrm{ms}(\tau_g)=0$

At higher orders ($n>1$), (near-)resonant $n$-photon processes of the form $\eta^{k+j}\Lambda_\Omega^{\prime \, n} \hat{J}_{\phi,y}^{n} \hat{a}^{\dagger k} \hat{a}^j$ give rise to parasitic spin-spin and spin-motion coupling.
For example, the leading effective contribution from two-photon processes to all orders is determined by $\vartheta(\tau)( e^{\eta^2}/{\eta^2})\hat{J}_{\phi,y}^2[\hat{\mathcal{D}}_{ 1}(\eta),\hat{\mathcal{D}}^\dagger_{1}(\eta)]=-\vartheta(\tau) \hat{J}_{\phi,y}^2(1 + 2 \eta^2 \hat{a}^\dagger \hat{a} +\mathcal{O}(\eta^4))$ where $\vartheta(\tau)$ is the first-order geometric phase in Eq.~\eqref{eq:ex:MS_Gate_eff_o2}.
With respect to slow ($\tau$-time) dynamics, the phonon-number-dependent contribution is second order in $\eta$ such that the gate is largely sensitive to the initial motional state.
Since the time-dependence appears only in the global prefactor, these contributions cannot be suppressed by applying a smooth, continuous pulse on the first sideband alone.
Suppression of higher-order processes requires, for example, additional beam tones addressing higher-order sidebands, providing further control over motional couplings \cite{sametiStrongcouplingQuantumLogic2021}.

Building on the insights from the QAT analysis, we pulse shape the coupling $\Omega$ and add additional frequency tones to suppress parasitic effects. 
In Fig.~\eqref{fig:MSgate_Full} (right), we adopt an experimentally accessible smooth window $\Lambda_\Omega(s) = \Lambda_\Omega \sin^4(\Lambda_\omega s/2)$ with a slow frequency $\Lambda_{\omega} =\lambda \Lambda_\xi \sim \mathcal{O}(\lambda)$.
This simple choice ensures that the first three orders of the dynamical phase vanish at the gate time $s_g=2\pi/\Lambda_\omega$, 
 corresponding to a single period of $\Lambda_\Omega(s)$.
To suppress leading-order out-of-Lamb-Dicke effects, we add a second frequency tone of the same form as in Eq.~\eqref{eq:ex:MS_Interaction_Hamiltonian} with coupling $\Omega_2$ and $\Delta_2=2\nu +\delta_2$ tuned near the second motional sideband.
For this pulse shape, we analytically find that $\delta_1=3\,\delta$, $\delta_2 = \delta$, $\delta = 3 \omega$, and $\Omega_2/\Omega_1 \approx0.7885$ constitute a sufficient set to avoid undesirable resonances with the window frequency and guarantee decoupling to third order \footnote{Ref. \cite{sametiStrongcouplingQuantumLogic2021} demonstrates a similar decoupling scheme sufficient to fourth order. Here, however, counter-rotating terms introduce motional dependence that is not suppressed.}.
The fourth-order QAT propagator,
$\hat{U}_I^{[4]}(s;\lambda) = \exp(-i \hat{\Phi}^{[3]}(s;\lambda)) \, \hat{U}_{\mathrm{eff}}^{[4]}(\tau;\lambda)\rvert_{\tau=\lambda s}$,
agrees closely with numerical integration in the strong coupling regime, achieving an average process fidelity $\mathcal{F}_\mathrm{avg} \simeq 1 - 10^{-5}$ over the gate time, consistent with the approximation's $\mathcal{O}(\lambda^4)$ error bound.
The inclusion of a smooth window strongly suppresses off-resonant terms such that $\hat{U}_{\mathrm{fast}}(s_g;\lambda) = e^{\mathcal{O}(\eta^5)} \approx \mathds{1}$. 
Despite strong coupling and an initial motional state with $\bar{n}=5$, for the given parameters we estimate an entangling gate fidelity of $\mathcal{F} = 0.9998(5)$.
This performance is comparable to state-of-the-art weak-field demonstrations \cite{ballanceHighFidelityQuantumLogic2016,gaeblerHighFidelityUniversalGate2016,clarkHighFidelityBellStatePreparation2021}.  
Notably, shaping the pulse with a smooth envelope contributes up to a $10\%$ improvement in the average gate fidelity relative to an unshaped pulse.

%% file: sections/fig_MS_level_diagram.tex
\begin{figure}[tp]
    \centering
    \includegraphics[width=0.99\columnwidth]{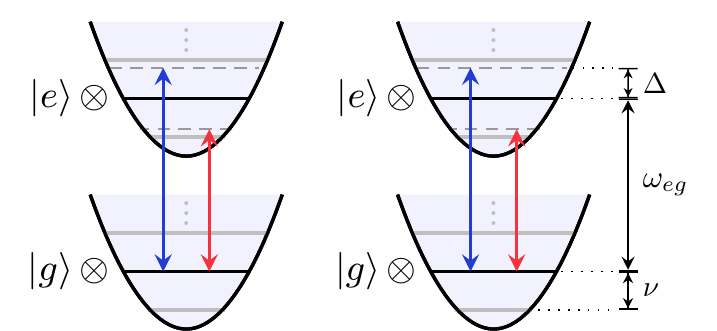}
    \caption{Level diagram: Two ion qubits coupled to a single phonon mode, symmetrically driven near the first motional sidebands by blue- ($+\Delta$) and red- ($-\Delta$) detuned drives.}
    \label{fig:level_diagrams}
\end{figure}

%% file: sections/fig_MS_Gate.tex
\begin{figure}[t!]
    \centering
    \includegraphics[width=\columnwidth]{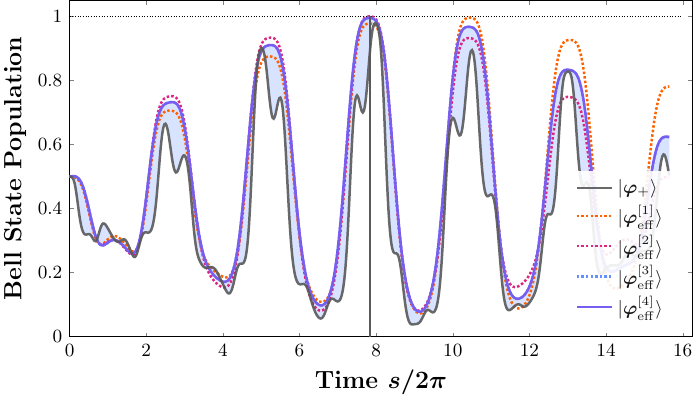}

    \caption{
    \textbf{Effective Hamiltonian dynamics} for population in the Bell state $\ket{\boldsymbol{\varphi}_{+}} = (\ket{ee} + \ket{gg})/\sqrt{2}$ of two ions under a strongly-coupled M{\o}lmer-S{\o}rensen gate in the carrier interaction picture are compared with numerical integration (dark gray), where $\ket{\bm{\varphi}_\mathrm{eff}^{[N]}}$ captures $N$th-order dynamics [eq.~\eqref{eq:qat:QAT_interaction_approx}]. 
    The system is initialized in $\ket{g g}$ with motional coherent state $\ket{\alpha=i\sqrt{5}}$ ($\bar{n}=5$). 
    Strong off-resonant and motionally-sensitive interactions shift the optimal entanglement time, and neglecting the carrier further reduces gate fidelity. 
    Nonetheless, the effective dynamics closely track the envelope of the numerical solution, with deviations highlighted by shading. 
    Parameters: $\eta = 0.1$, $\Omega = \nu$, $\phi_+ = \pi/4$, $\phi_- = 0$, and $\delta \approx 0.383\nu$.
    }
    \label{fig:MSgate_Eff}
\end{figure}

\begin{figure*}[t!]
    \centering
    \includegraphics[width=\textwidth]{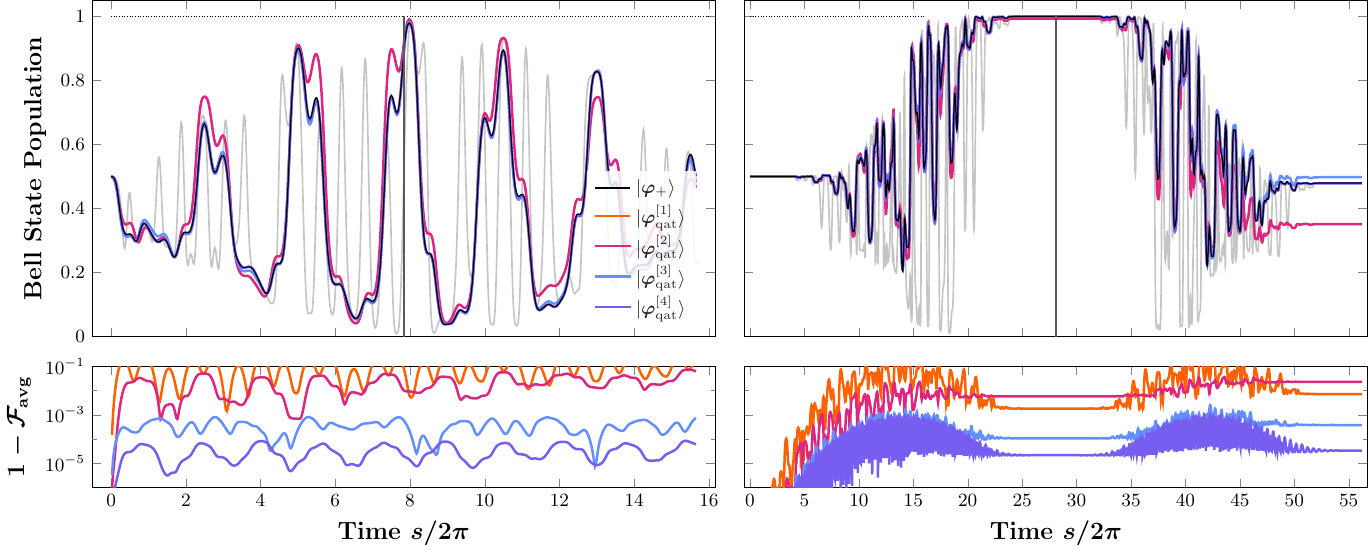}

\caption{
\textbf{Full QAT dynamics} for population in the Bell state $\ket{\boldsymbol{\varphi}_{+}} = (\ket{ee} + \ket{gg})/\sqrt{2}$ of two ions under a strongly-coupled M{\o}lmer-S{\o}rensen gate in the carrier interaction picture as compared to numerical integration (black). The dynamics including the off-resonant carrier is shown for reference (light gray).
The system is initialized in $\ket{g g}$ with a motional coherent state $\ket{\alpha=i\sqrt{5}}$ ($\bar{n}=5$). 
\textbf{Left}: includes off-resonant interactions neglected in the effective Hamiltonian description of Fig.~\eqref{fig:MSgate_Eff}. 
\textbf{Right}: By adding pulse shaping, a power-of-sine window suppresses off-resonant effects, while a second-sideband tone decouples from leading-order thermal effects; parameters: $\Omega_2/\Omega_1 \approx0.7885$, $\delta_1/3=\delta_2 \approx 0.107\nu$, $\omega=\delta_2/3$, $s_g = 2\pi/\Lambda_\omega$.
\textbf{Bottom}: shows the rapid convergence of the QAT approximation to the numerical result via instantaneous process fidelity, \textit{i.e.}, the average fidelity between numerical and approximate dynamics over a Pauli eigenstate 2-design \cite{nielsenQuantumComputationQuantum2010,nielsenSimpleFormulaAverage2002a}.
}
    \label{fig:MSgate_Full}
\end{figure*}

%% file: sections/V_Conclusion.tex
In summary, we presented a unitarity-preserving, multi-timescale QAT framework for analytically studying gate performance from slow- and fast-varying interactions.
Notably, our approach distinguishes itself by retaining and accounting for the details of the off-resonant, out-of-RWA dynamics to arbitrary order of precision.
We demonstrated the robust characterization of gate dynamics and fidelity under strong spin-motion coupling for the fast-entangling M{\o}lmer-S{\o}rensen gate.
For this example and others, the advantage of a scalable, analytic, and unitary framework is twofold: it provides physical insight by isolating the desired gate interaction while revealing unwanted couplings, and it reduces the computational cost of simulating large systems, such as infinite-dimensional bosonic modes that are prone to truncation errors. 
By retaining off-resonant processes, QAT offers a transparent way to assess coherent leakage channels and parameter sensitivity, supporting gate robustness analysis beyond the reach of standard RWA treatments.
The framework is also naturally compatible with numerical optimal control methods \cite{khanejaOptimalControlCoupled2005,motzoiSimplePulsesElimination2009,machnesComparingOptimizingBenchmarking2011,reichMonotonicallyConvergentOptimization2012,glaserTrainingSchrodingersCat2015}, providing reduced effective models that can serve as analytic starting points or efficient seeds for computationally intensive pulse optimization in systems with strong, off-resonant effects.
By separating fast and slow timescales, QAT enables improved identification of coherent error sources and efficient simulation, thereby informing enhanced error suppression strategies for high-fidelity quantum computation.

\begin{acknowledgments}
    We gratefully acknowledge the late Alex Levine for his early guidance and inspiring discussions.
    We also thank Robijn Bruinsma for valuable comments on the manuscript, and Alexander Radcliffe and Ken Brown for fruitful conversations on the Mølmer–Sørensen interaction in trapped-ion architectures.
    This work was supported by the National Science Foundation under Grants PHY-2207985 and OMA-2016245, and by the U.S. Army Research Office under Grant W911NF-24-S-0004.
\end{acknowledgments}

\subsection*{CRediT Author Statement}

\textbf{Kristian D. Barajas:} Conceptualization, Investigation, Formal analysis, Writing – Original Draft, Writing – Review \& Editing 

\textbf{Wesley C. Campbell:} Supervision, Validation, Writing – Review \& Editing, Funding Acquisition

%% file: supplemental_arxiv.tex
\pagebreak
\widetext

\begin{center}
\textbf{Supplemental Material for\\
``\textit{Quantum gate dynamics beyond the rotating-wave approximation using multi-timescale quantum averaging theory}''}
\end{center}

\setcounter{equation}{0}
\setcounter{figure}{0}
\setcounter{table}{0}
\setcounter{page}{1}
\makeatletter
\renewcommand{\theequation}{S\arabic{equation}}
\renewcommand{\thefigure}{S\arabic{figure}}
\renewcommand{\bibnumfmt}[1]{[S#1]}
\renewcommand{\citenumfont}[1]{S#1}

This Supplemental Material provides additional derivations and technical details that support the results presented in the main text.
All equation and figure references to the main manuscript are
denoted explicitly.
We emphasize that the results presented here do not introduce new physical claims beyond those of the main article but instead provide additional details to support the reproducibility of the main results.

\section{Formal Series Expansion of Displacement Operator}\label{appendix:displacement_operator}
The formal exponential power series of the displacement operator in powers of $\eta$ returns
\begin{equation}
    \hat{\mathcal{D}}(\eta;\theta) = \sum_{n=0}^\infty  \frac{(i \eta)^n}{n!} (\hat{a}^\dagger(\theta) + \hat{a}(\theta))^n.
\end{equation}
However, this expansion is not in normal-ordered form.
Instead, we expand the time-periodic displacement operator in a Fourier series with $\theta=\Lambda_\nu s$ yielding
\begin{equation}
\begin{aligned}
        \hat{\mathcal{D}}(\eta;\theta) ={}& e^{i \eta (\hat{a}^\dagger(\theta) + \hat{a}(\theta))}, \quad \hat{a}^\dagger(\theta) = \hat{a}^\dagger e^{i \theta}  \\
        ={}& \sum_{k \in \mathbb{Z}} \hat{\mathcal{D}}_k(\eta) e^{i k \theta}
\end{aligned}
\end{equation}
where
\begin{equation}
    \begin{aligned}\label{eq:appendix:displacement_op:Fourier}
        \hat{\mathcal{D}}_k(\eta) ={}& \frac{1}{2\pi} \int_{-\pi}^{\pi} e^{i \eta (\hat{a}^\dagger(\theta) + \hat{a}(\theta))} e^{-i k \theta} d\theta \\
        ={}& e^{-\eta^2/2} \sum_{\mathclap{n=\frac{|k|-k}{2}}}^\infty \; \frac{(i \eta)^{2n+k}}{(n+k)! n!} \hat{a}^{\dagger n+k} \hat{a}^n \\
    \end{aligned}
\end{equation}
where $\mathcal{\hat{D}}_{k}(i\eta)$ is a Bessel-Clifford operator satisfying $\mathcal{\hat{D}}_{-k}(i\eta) = \hat{\mathcal{D}}_k^{\dagger}(-i \eta) = (-1)^k \mathcal{\hat{D}}_{k}^{\dagger}(i\eta)$.
From the normal-ordered form in Eq.~\eqref{eq:appendix:displacement_op:Fourier}, the displacement operator can be expressed in powers of $\eta$ as
\begin{equation}
    \mathcal{D}(\eta;\theta) = e^{-\eta^2/2}\sum_{n=0}^\infty (i\eta)^n \, \hat{\mathcal{D}}^{(n)}(\theta)
\end{equation}
where
\begin{equation}\label{eq:appendix:displacement_op}
\begin{aligned}
    \mathcal{\hat{D}}^{(n)}(\theta) ={}& \sum_{k=0}^n \frac{a^{\dagger n-k} a^k}{(n-k)!k!}e^{i (n-2k) \theta}
\end{aligned}
\end{equation}
is the $n$th Taylor polynomial.

\section{QAT Results for MS Gate Dynamics}\label{appendix:MS_Gate_DynamicalPhase}
We provide the order-by-order effective Hamiltonian and dynamical phase contributions for the M{\o}lmer-S{\o}rensen interaction Hamiltonian in Eq.~(13), derived using QAT and shown in Fig.~(2) and (3) (left) in the main text.
The first-order terms capture the leading spin-motion coupling, while higher-order terms account for off-resonant effects, motionally sensitive interactions, and nonlinear corrections. 
Third- and fourth-order contributions become increasingly important for strong driving or large initial motional amplitudes, and are essential for accurately predicting gate fidelity and timing beyond the lowest-order approximation. 
The effective Hamiltonian contributions to fourth order:
\begin{subequations}
    \begin{align}
        \lambda \hat{H}^{(1)}_{I,\mathrm{eff}} ={}& - \eta \Lambda_{\Omega}'  \hat{J}_{\phi,y} (\hat{a}^\dagger G(\tau) + h.c.) \\
        \lambda^2 \hat{H}_{I,\mathrm{eff}}^{(2)} ={}& \eta^2 \Lambda_\Omega^{\prime 2} \frac{1}{\Lambda_{\delta}-2\Lambda_\nu} \hat{J}_{\phi,y}^2 \\ 
        \lambda^3 \hat{H}_{I,\mathrm{eff}}^{(3)} ={}& \frac{1}{2} \eta^3 \Lambda_\Omega' \hat{J}_{\phi,y}(\hat{a}^{\dagger 2} \hat{a} \, G(\tau) + h.c.) \\
        \begin{split}
        \lambda^4 \hat{H}_{I,\mathrm{eff}}^{(4)} ={}& \eta^4 \Lambda_\Omega^{\prime 2} \hat{J}_{\phi,y}^2\Bigr[ - \frac{2}{\Lambda_\delta-2\Lambda_\nu} \hat{a}^\dagger \hat{a} \\
        & +\frac{4\Lambda_\nu}{(\Lambda_\delta -3\Lambda_\nu)(\Lambda_\delta+\Lambda_\nu)}  (\hat{a}^\dagger \hat{a} + 1/2) \\
        &+ \Bigl(\frac{\Lambda_\nu(5\Lambda_\nu^2-2\Lambda_\delta^2)G(\tau)^2}{(\Lambda_\delta^2 -\Lambda_\nu^2)(\Lambda_\delta^2 -(2\Lambda_\nu)^2)}  \hat{a}^{\dagger 2} + h.c. \Bigr) \Bigr]
        \end{split}
    \end{align}
\end{subequations}
and for the dynamical phase:
\begin{subequations}
    \begin{align}
        \lambda \hat{\Phi}^{(1)} =& -i \frac{\eta\Lambda_\Omega'}{\Lambda_\delta-2\Lambda_\nu}G^*(s)e^{-2i\Lambda_\nu s} \hat{J}_{\phi,y} \hat{a}^
    \dagger + h.c. \\
    \begin{split}
    \lambda^2 \hat{\Phi}^{(2)} =& -\frac{2 \eta ^2 \Lambda_\Omega' }{\Lambda_\delta -\Lambda_\nu}\cos ((\Lambda_\delta-\Lambda_\nu )s + \phi_{-}) \hat{J}_{\phi,y} \hat{a}^\dagger \hat{a} \\
    & -\Bigl(\frac{1}{2} \eta ^2 \Lambda_\Omega'  \Bigl(\frac{G(s)e^{i s\Lambda_\nu }}{\Lambda_\delta +\Lambda_\nu } \\
    &\quad+\frac{G^*(s)e^{3i \Lambda_\nu}}{\Lambda_\delta -3 \Lambda_\nu }\Bigr)  \hat{J}_{\phi,y} \hat{a}^{\dagger 2} + h.c.\Bigr) \\
    &+\frac{\eta^2 \Lambda_\Omega^{\prime 2} }{(\Lambda_\delta -\Lambda_\nu)(\Lambda_\delta -2\Lambda_\nu)} \\
    &\qquad \times \sin (2((\Lambda_\delta-\Lambda_\nu )s + \phi_{-})) \hat{J}_{\phi,y}^2
    \end{split} \\
    \begin{split}
        \lambda^3 \hat{\Phi}^{(3)} ={}& \eta ^3 \Lambda_\Omega'\Bigr[ \frac{i G^*(s)e^{i s 2 \Lambda_\nu }}{2 (\Lambda_\delta -2 \Lambda_\nu )} \hat{J}_{\phi,y} \hat{a}^{\dagger 2} \hat{a} \\
        +& i\Bigl(\frac{ G^*(s) e^{i s 4 \Lambda_\nu }}{6 (\Lambda_\delta -4 \Lambda_\nu)}\\
        &\qquad -\frac{ G(s)e^{i s 2 \Lambda_\nu }}{6 (\Lambda_\delta +2 \Lambda_\nu)} \Bigr) \hat{J}_{\phi,y} \hat{a}^{\dagger 3} \\
        +& \frac{\Lambda_\Omega'}{2}\Bigl(\frac{ \left(2 \Lambda_\delta^2+\Lambda_\delta  \Lambda_\nu -7 \Lambda_\nu ^2\right) G(s)^2 e^{-i s \Lambda_\nu}}{(\Lambda_\delta -2 \Lambda_\nu ) (\Lambda_\delta^2 -\Lambda_\nu^2 ) (2 \Lambda_\delta -\Lambda_\nu) } \\
        +&\frac{\left(2\Lambda_\delta ^2-5 \Lambda_\delta  \Lambda_\nu +\Lambda_\nu ^2\right) G^*(s)^2 e^{i s 3 \Lambda_\nu}}{(2 \Lambda_\delta-3 \Lambda_\nu )(\Lambda_\delta -3 \Lambda_\nu ) (\Lambda_\delta -2 \Lambda_\nu )(\Lambda_\delta -\Lambda_\nu )} \\
        +&  \frac{2(\Lambda_\delta -7\Lambda_\nu )e^{i \Lambda_\nu  t}}{(\Lambda_\delta -3\Lambda_\nu ) (\Lambda_\delta^2 -\Lambda_\nu^2 )}  \Bigr) \hat{J}_{\phi,y}^2 \hat{a}^\dagger \Bigr] +h.c.
    \end{split}
    \end{align} 
\end{subequations}